\documentclass[12pt]{article}

\begin{document}

\title{\bf ATOMIC PHYSICS: COMPUTER CALCULATIONS AND THEORETICAL
ANALYSIS~}
\author {E. G. Drukarev\\
{\em Petersburg Nuclear Physics Institute,}\\
{\em St. Petersburg, Gatchina 188300, Russia}}
 \date{}

\maketitle

\begin{abstract}
It is demonstrated, how the theoretical analysis preceding the numerical
calculations helps to calculate the energy of the ground state
of helium atom, and enables to avoid qualitative errors in calculations
of the characteristics of double photoionization.  \end{abstract}

\section{Introduction}

As it stands now, many of the publications devoted to interaction of  photons
and electrons with atoms and molecules have similar structure. In the first
step a most general quantum mechanics formula for the cross section is
written. In the next step the most "accurate" numerical functions for the
bound state electrons are found. Since the binding energy can be determined
with high accuracy, the wave functions pass the test for reproducing the
binding energies with a good accuracy. In the next step computer is at work.
Sometimes there is a possibility to try the influence of the final state
interactions. The usual choices are: plane waves, Coulomb functions...

About twenty years ago such approach was justified to  some
 extent. Most
calculations were connected with the characteristics, which could be
detected in the experiments. In the real experiments the targets
obtained the energies of the order of the electron binding energies.
There was no small parameter, and thus no possibility to evaluate the
equations, determined by  the original formalism of quantum
mechanics. The part of atomic physics, related to the real experiments was
indeed becoming the science about the precise computations. The near-threshold
behavior of the processes, where one could find a small parameter, was rather
an exception.

The situation changed at the end of 80-th, when new synchrotron sources of
the photons became available. The experiments with the photon energies of
about 10 $keV$ went on. For the light atoms these energies  are much
larger then the binding energies. A small parameter emerged in the theoretical
problems, which could be of practical importance.

The general approach to the "high energy atomic physics" can be based on the
works of Bethe \cite{1}. A somewhat modernized presentation
is given in \cite{2,3}. Being initially formulated
 for the electron-atomic scattering, these principles can be expanded
 to any process of the interaction between the high energy particles with
 the bound systems. The wording "high energy" means that the energy exceeds
 strongly the binding energy of the system. The cross-sections of such
 processes
 can be expressed through certain parameters of the bound systems. The
 interactions between the fast and slow participants can be treated as perturbation.
 Each act of transferring of large momentum leads to a parametrically
 small factor. In the region, where the processes are kinematically allowed
 for the free electrons, a small momentum $q$  of the order of the average
 momentum of the bound electron
 $\eta$ is transferred to the nucleus. The processes outside this
kinematical region
 ( which we shall refer to as {\em Bethe surface}) require  large momentum
 $q \gg \eta$ to be transferred to the nucleus. The processes outside the
 Bethe surface are thus strongly quenched.

 Hence, the amplitudes  of the high energy atomic physics processes depend on
 the  parameter
\begin{equation}
 \kappa\ =\ \eta/q\ .
  \end{equation}
  On the Bethe surface $\kappa$ is of the order of unity, while
  $\kappa \ll 1$ outside the Bethe surface. Using the system of units
  with $\hbar=c=1$, we can write for the lowest state
  of the hydrogenlike atom
 \begin{equation}
 \eta\ =\ m\alpha Z\ ,
  \end{equation}
   with $m$ being the electron mass, $\alpha=1/137$ is the fine
   structure constant, while Z is the charge of the nucleus. We shall use
   this relation for estimations in the many-electron cases as
well.

   The
perturbative treatment of the fast electrons, which is another point
   of the Bethe theory, leads to certain misunderstandings sometimes.
   The sections "Perturbation theory" of the textbooks on quantum mechanics
   often treat only the short-ranged forces. Indeed,
   iteration series term of the Lippman-
   Schwinger equation for the electron wave function in the Coulomb field
   of the nucleus contains the infrared-divergent terms. In this sense,
   there is no such thing as "perturbative series
   for the Coulomb field" and
   the words " the fast electrons can be described by the plane waves " become
   confusing. However, the solution
   was found long ago. It was guessed by Dalitz \cite{4} and proved  by
   Gorshkov \cite{5}, that the divergent terms compose the factor $exp(i\Phi)$
   with real $\Phi$ in the amplitudes, thus cancelling in the
cross-sections.  Hence a way to solve the problem is to present the
electromagnetic field of the nucleus in the form $V(r)=-\alpha Z
e^{-\lambda r}/r$ with $\lambda \rightarrow 0$.  Now the Lippman-
   Schwinger equation and the perturbative theory become legitimate,
   while the terms which depend on $\lambda$ (i.e. $ln\lambda$) cancel
   in the cross-sections.

   The perturbation parameter of the nonrelativistic electron with the
   three-dimensional momentum $p$, moving
   in the Coulomb field
   is \cite{6}
     \begin{equation}
 \xi_Z\ =\ m\alpha Z/p\ .
  \end{equation}
  The condition $\xi_Z \ll 1$ means that the relative contribution
  provided by this interaction to the electron wave
  function is much smaller than unity \cite{6}. Since different regions
  of distances may be important for different terms of the expansion
  of the wave function, one should be careful and expand the
  {\em amplitude} in powers of $\xi_Z$. The small distances
  are connected with transferring of large
  momentum $q\gg\eta$ to the nucleus. This process can be considered
  as perturbation as well, with $\kappa^4 \ll 1$ being the parameter of
  expansion-Eq(1).

  However, one often finds large deviations between the calculations with the
  plane waves and Coulomb field functions applied for the description of
  the continuum electrons, even at large values of energies. Sometimes
   this leads to the statement that "plane wave approximation never
  works".  However, as explained in \cite{7}, there are two parameters,
  which depend on the electron momentum $p$. Besides the parameter
  $\xi_Z^2$ which is the true parameter of the expansion of the wave
  function at the distances of the order of the
size of the atom, there is a parameter $\pi \xi_Z$,
  which enters through the normalization factor of the Coulomb wave function
  anyway. In the processes with the large momentum  transferred to the
  nucleus, there are some more terms, containing the parameter $\pi \xi_Z$.
  Fortunately, they also compose a factor. Thus, assuming
  \begin{equation}
 \pi\xi_Z \approx 1 ,~~~~~~~~~~~~~~ \xi_Z^2 \ll 1
  \end{equation}
  one can carry out expansion in powers of $\xi_Z^2$. Also, the factors
  depending on the parameter $\pi \xi_Z$ cancel in the ratios of the cross
  sections of the processes containing the high energy electrons with the
  close values of energies. The recent calculations, presented in \cite{8}
  illustrated the general validity of the latter statement.

  The attempts to include the final state interactions of the outgoing
  electrons in a straightforward way also met some
  difficulties, caused by the long-range nature of the interaction. The
  approach which enabled to include the $e-e$ interaction in the lowest
  order of the   parameter of their interaction
  \begin{equation}
 \xi^2\ =\ (m\alpha)^2/p_r^2,
  \end{equation}
with $p_r$ being the linear momentum of the relative motion
was worked out in \cite{9}.

  Since the high-energy experiments were not available in atomic physics for
  a long time, these ideas were not applied much. The only group which
  systematically used and developed the approach was that of V. G. Gorshkov
  and his collaborators \cite{10}, who focused on the Coulomb field
calculations.

  Now one can conclude  that  it is not enough to require that the bound state
  wave function provides an accurate energy value. It should provide the
  adequate description of the electron density on the distances,
  which are important  for the considered process. If the process is not
  allowed for the free electrons, the important distances can be of minor
  importance for determination of the binding energy. In treating the final
  states one can use the perturbative expansions in powers of $\xi_Z^2$,
  $\xi^2$ and $\kappa^2$, taking care that all the terms of the leading
  order are picked in the amplitude.

  It looks to me, that the story of investigation of the double photoionization
  of helium provides good illustrations of how the application of this approach
  enabled to obtain the results. It illustrates also, how ignoring of these
  ideas led to the erroneous results. The theory gave two lessons: at the end
  of 60-th and at the end of 90-th. The relativistic aspects of the
 problems will not be touched. I
  will show , how the main results can be obtained by the standard
  methods of nonrelativistic quantum mechanics. I will show also, that
 the Lippman- Schwinger equation appears to be a powerful tool.
 However, I start with the example of a static problem of the
 calculation of the ground state binding energy of the helium atom.

\section{Calculation of the ground state energy of  helium atom}

The binding energy of the ground state of helium atom is known from
the experiments with very high precision. The
relative accuracy was $10^{-6}$
in  late 50-th \cite{2}.
It is about $2 \cdot 10^{-7}$ today \cite{11}.
This stimulated the attempts
to solve the Schr\"odinger equation for the helium ground state as
 precisely as possible.

The Hartree-Fock (HF) method, developed in early 30-th,
reproduces the value of the binding energy
with the relative error of about $1.5\%$. The experiments are much more
precise. The reasons of this failure are known. The HF function does not
depend on the distance between the electrons. The probability to find the
two electrons to be close to each other is thus overestimated in HF approach.

An alternative approach was developed in the same years as the HF method.
It consists in employing approximate wave functions
$\psi_a(r_1,r_2, r_{12})$,
presented by explicit analytical formulas. Here $r_1,r_2$ denote the
distances between the electron and the nucleus, $r_{12}$ is the distance
between the electrons. The lower index $a$ stands for "approximate".
These functions contain a number of parameters. The values of the
parameters are chosen to minimize the expectation value
\begin{equation}
E\ =\ \langle\psi_a|H|\psi_a\rangle\ ,
\end{equation}
of the Hamiltonian $H$ of the helium atom. The simplest example is the
product of two hydrogenlike functions in the field of the nucleus
with certain "effective nuclear charge" $Z_e$. The effective charge plays
the role
of the variation parameter. In this case
\begin{equation}
\psi_a(r_1,r_2)=\psi_h(r_1)\psi_h(r_2),
\end{equation}
with the well known analytic expression for $\psi_h(r)=e^{-ar}$,
$a=m\alpha Z_e$. ( All the wave functions are
normalized to $\psi(0)=1$).The solution $Z_e=27/16$ \cite{6} provides the
energy value with the error of $2\%$.

 More complicated functions of the form
\begin{equation}
\psi_a(r_1,r_2)\ =\ e^{-a(r_1+r_2)}P(r_1,r_2,r_{12})
\end{equation}
with $a$ and the coefficients of the polynomial
\begin{equation}
P(r_1,r_2,r_{12})=\Sigma c_{ijk}r_1^ir_2^jr_{12}^k
\end{equation}
being the variation parameters, have been used since late 20-th.
The earlier versions contained 3 parameters. By the end of 50-th the
functions with more than 200 parameters were employed \cite{2}.
The functions $P$, containing also the dependence on the ratios
 $\frac{r_{12}}{(r_1+r_2)}$ and $\frac{(r_1-r_2)^2}{r_{12}^2}$ have been
used. This enabled to obtain the convergence of the results
to a certain value, interpreted as the true energy value,
with the precision of several units of $10^{-10}$, reproducing
5 signs in the value of the binding energy ( The calculations
with the accuracy better then $10^{-4}$ require the account of
the finite mass and motion of the nucleus, inclusion of
relativistic corrections, etc).

However, Barlett {\em et al.} \cite{12} showed that the functions (8),
with $P$ having a polynomial structure could not satisfy the
Schr\"odinger equation for the ground state of helium. Such solution
does not exist since it does not have a proper limit at the triple
coalescence point $r_1=r_2=r_{12}=0$. The functions which treated the
three-particle singularity of the Hamiltonian were found by Fock
\cite{13}. They took the form \begin{equation} P(r_1,r_2,r_{12})=\Sigma
c_{ijkn}r_1^ir_2^jr_{12}^k[ln(r_1^2+r_2)^2]^n, \end{equation} thus
containing the logarithmic terms. (Usually, the Fock function is
presented by using the hyperspherical variables \cite{14}).

Inclusion of the logarithmic terms has little effect on the energy value.
 However, account of these terms improves
 the convergence of the
variation calculations. Also, inclusion of such terms enables to diminish
the number of variation parameters. For example, the calculation with
the accuracy $10^{-9}$ required 1078 parameters of the function (9),
while it was sufficient to use the function (10) with 52 parameters
to achieve the same precision \cite{15}. Furthermore, as noted by
Myers {\em et al.} \cite{16}, for some of the functions (9)
 the variation procedure chooses the parameters in such a way, that
the expansion mimicked the function (10) with a smaller number of
parameters.

Behavior of the true ground state wave function $\Psi(r_1,r_2,r_{12})$
at the two-particle coalescence points is determined by the Kato cusp
conditions \cite{17}. In the point of electron-nucleus coalescence
$r_1=0$ it is
\begin{equation}
r_0\frac{\partial{\Psi(r_1,r_2,r_2)}}{\partial{r_1}}=-Z\Psi(0,r_2,r_2),
\end{equation}
while in the electron-electron coalescence point $r_{12}=0$ it is
\begin{equation}
r_0\frac{\partial{\Psi(r_1,r_1,r_{12})}}{\partial{r_{12}}}=\frac{1}{2}
\Psi(r_1,r_1,0),
\end{equation}
with $r_0=\frac{1}{m\alpha}$ standing for the Bohr radius.
Equations (11,12) can be
viewed as the conditions of the cancellation of the singular
terms in the Schr\"odinger equation.

 Let us illustrate this, considering the
single-electron wave function $\psi(r)$ of the
bound electron in $1s$
state of certain effective
field $U(r)$, which approximates the
interactions with the nucleus and with the other electrons.
The Schr\"odinger equation for this case is
\begin{equation}
\frac{-1}{2m}\left(\psi^{''}(r)+\frac2r\psi'(r)\right)+U(r)\psi(r)
=E\psi(r).
\end{equation}
 At $r\rightarrow 0$ the interaction $U(r)$ is determined
by interaction with the nucleus. Thus, $U(r \rightarrow 0) \sim
-\frac{\alpha Z}{r}$. Hence, for $ r\rightarrow 0$ we can write
\begin{equation}
\frac{-1}{2m}\psi^{''}(r)+\lim_{r\to0}\left(\frac{\frac{-1}m\psi'(r)}r
-\frac{\alpha Z\psi(r)}{r}\right)=E\psi(r).
\end{equation}
Since both sides of Eq. (14) should be regular, the expression in
brackets should have a finite value.
Thus
\begin{equation}
r_0\frac{\partial{\psi(r)}}{\partial{r}}\ =\ -Z\psi(0)\ ,
\end{equation}
at $r\rightarrow0.$  Similar analysis of the
Schr\"odinger equation for the atom of helium
leads to Eqs.(11),(12). Note, that these relations could have been
obtained immediately after the Schr\"odinger equation  was written in
1926. Surprisingly, they were found only thirty years later.

The proper treatment of the two-particle singularity is even
more important than the account of the three-particle one \cite{16}.
This is because the three particle coalescence point is
felt in the smaller volume of integration of the matrix element (6).
Myers {\em et al.} \cite{16} suggested the simplest form of the wave
function, which satisfies both Kato conditions
\begin{equation}
\psi_a(r_1,r_2, r_{12})\ =\ e^{-a(r_1+r_2)+br_{12}}
\end{equation}
with $a=m\alpha Z$, $b=\frac{m\alpha}{2}$. It is amusing that this function,
which does not contain fitting parameters , provides the same precision in
determination of the binding energy, as the function (7), which contains a
variation parameter $Z_e$. The function (16) knows nothing about the
three-particle singularity. The simplest function, which would account
both two- and three-particle singularities is described by a more complicated
formula, also presented in $\cite{16}$.

If the approximate wave function is chosen in a proper form, the variation
procedure exhibits the tendency to find the parameters, providing small errors
in the Kato conditions (11,12). The wave functions
\begin{equation}
\psi_a(r_1,r_2, r_{12})= e^{-a(r_1+r_2)+br_{12}}
\Sigma c_{ijk}r_1^ir_2^jr_{12}^k,
\end{equation}
($i+j+k \leq N$) analyzed by Teng and Shakeshaft \cite{18} provide the
relative errors of about $6\%$ and $25\%$ in Eqs.(11, 12) correspondingly,
even in the
case of 4 parameters (N=1). Although the Kato conditions are satisfied only
approximately, it helps in improving the convergence. The function (17)
 with 14 parameters (N=4) provides the value of the binding energy with
the relative deviation $3 \cdot 10^{-5}$ from the one, obtained with
the functions, described by Eqs (8,9) and containing 210 parameters
\cite{2}.

In the nowadays studies the subject of investigation is usually not
only the expectation value (6), but also the local energy functional
\begin{equation}
E(\vec r_1, \vec r_2)\ =\  \frac{H\psi_a}{\psi_a}\ ,
\end{equation}
which gives more detailed information about the relation of the
approximate function $\psi_a$ to the true wave function $\Psi$.
Another improvement consists in developing the approaches, based on direct
solution of the Schr\"odinger equation. In one of the methods, suggested
by Haftel and Mandelzweig ( see \cite{19}, where the other methods are
also discussed) the wave function is presented as a
product of a correlation factor, describing the coalescence points
singularities
and a smooth function, expanded in a series of hyperspherical harmonic
functions. The detailed description of these points is
beyond the scope of this lecture.

I hope that this Section made one convinced that taking care of the proper
analytical structure of the approximate wave functions, one obtains
the better results. Now we shall see that this becomes increasingly important
for the dynamical problems.

\section{Asymptotics  of the double photoionization cross section:
$\omega^{-7/2}$ or $\omega^{-5/2}$~?}

The double photoionization of the ground state of
helium is studied since the late 50-th.
The process provides a direct probe of three-body physics, since the
three charged particles are involved. A huge number of experimental results
gives a good possibility for theoreticians to test their approaches.

The mechanism, which determines the main contribution to the
nonrelativistic
asymptotics of the cross section $\sigma^{2+}(\omega)$ with $\omega$
standing for the photon energy, was clarified by Kabir and Salpiter
\cite{20}. At the values of $\omega$, exceeding strongly
the binding energy $I$, the process can be viewed as a single
photoionization, followed by the transition of the second electron
to continuum. The latter process takes place due to the change of the
Hamiltonian of the system. This is the analog of the shake-off
({\em vstryahivanie}), suggested by Feinberg \cite{21} for the
description of ionization during the nuclear beta decay, due to the
sudden change of the electromagnetic field, caused by the change of the
charge of the nucleus. However, using the same wording {\em shake-off},
one must remember, that, in contrast to the beta decay, the mechanism
can not be treated as due to the sudden change of certain effective
{\em field}.  Byron and Joachain \cite{22} showed that, being
attributed to the sudden change of self-consistent (Hartree-Fock)
field, the mechanism underestimates the asymptotic value by a factor of
$3$.  The initial state correlations beyond the effective field are
thus of crucial importance.

The electron, which interacts with the photon directly, absorbs most of
the energy $\epsilon_1 \approx \omega$. The secondary electron carries the
energy of the order of the binding energy $\epsilon_2 \sim I \ll \omega$.
Thus, the energy dependence of the amplitude and of the cross section
of the high energy nonrelativistic photoionization are the same, as that
for the single photoionization process.

Recall now the mechanism of single ionization.

\subsection{Asymptotics of the single photoionization}

Here we assume that the electrons are described by the single-particle
wave functions. The general expression for the amplitude of ionization of
$1s$ state is
\begin{equation}
F^+= (4\pi \alpha)^{1/2}\int{d^3r \psi^*_{p1}(r)\gamma e^{i(kr)}\psi(r)},
\end{equation}
with $\psi_{p1}$ being the continuum wave function with the asymptotic
momentum $\vec p_1$, $\psi$ describes the bound $1s$ electron, while
\begin{equation}
\gamma=-i\frac{(\vec e ~\vec \nabla_r)}{m},
\end{equation}
is the operator of interaction between the photon and electron, $\vec e$
stands for the photon polarization. The Fourier transform provides
\begin{equation}
F^+= (4\pi \alpha)^{1/2}\int{\frac{d^3q}{(2\pi)^3} \psi^*_{Fp1}(q)
\frac{(eq)}{m}\psi_F(q-k)},
\end{equation}
Here the functions $\psi_F$ are the Fourier transforms of the functions
$\psi$. The process requires that momentum
\begin{equation}
\vec Q\ =\ \vec k-\vec p_1\
\end{equation}
is transferred to the recoil ion. Note, that the energy of the fast electron
$\epsilon_1=p_1^2/2m \gg I$, while the binding energy $I=\frac{\eta^2}{2m}$,
with $\eta$ standing for the average momentum of the bound electron. Thus,
$\epsilon_1 \gg I$ means, that $p_1 \gg \eta$, i.e. the momentum of the fast
electron is much larger than that of the bound electron. Also, the energy
dependence of the photon and electron momenta are $k=\omega$ and
 $p_1 \approx
(2m\omega)^{1/2}$. Hence, $Q \gg \eta$ in this process.
The large momentum (i.e. momentum, exceeding strongly the average
value for
 the bound electron) should be transferred to the recoil ion.

Start with the plane wave
\begin{equation}
\psi_{Fp1}(\vec q) =\psi^{(0)}_{Fp1}(\vec q)=
(2\pi)^3\delta(\vec p_1-\vec q),
\end{equation}
for the description of the final state electron. In this case all
momentum $Q$ is transferred to the recoil ion by the initial state electron.
The contribution to the amplitude is
\begin{equation}
F^+_0=\ (4\pi\alpha)^{1/2}\frac{(ep_1)}m \psi_F(p_1)\ .
\end{equation}
For the hydrogenlike functions the asymptotics $\psi_{hF}(p_1)=\frac{8\pi^{-1/2}
\eta^{5/2}}{p_1^4}+O(p_1^{-6})$ at $p_1 \gg \eta$ can be obtained
explicitly. One can expect the $p_1^{-4}$ behavior
to be true in the general case,
since the momenta $p_1 \gg \eta$ correspond to small distances
$r \ll \eta^{-1}$, where an  electron function $\psi$ has a hydrogenlike
shape. Anyway, this can be proved rigorously, and the pre-asymptotic
coefficient can be calculated. Since the $s$ state wave function
does not depend on the direction of the vector $\vec r$,
we can evaluate
\begin{equation}
\psi_F(p)=\int{d^3r \psi(r) e^{-i(\vec p \vec r)}}=
-\frac{4\pi}{p^2}\int{dr r \psi(r)[cos(p r)]'}=...
-\frac{8\pi \psi'(0)}{p^4}+ O(p^{-6})
\end{equation}
Here the last equality is obtained after using two integrations
by parts.

To get convinced, that Eq.(25) indeed presents the asymptotics of the
amplitude, let us try to include the interactions between the outgoing
electron and the recoil ion. The lowest order
perturbative correction to the outgoing electron
wave function, caused by interaction with the nucleus $V^C$can be obtained by
using the standard formula of quantum mechanics
\begin{equation}
\psi_n^{(1)}=\Sigma\frac{V^C_{nm}\psi^{(0)}_m}{\epsilon_n-\epsilon_m},
\end{equation}
with the sum over the states $m$, which do not coincides with $n$,
$\epsilon_{n,m}$ stand for the unperturbed energy values.
This provides
\begin{equation}
\psi^{(1)}_{Fp1}(\vec q)=-\frac{8\pi\alpha Z m}{(\vec p_1-\vec q)^2
(p_1^2-q^2)},
\end{equation}
with the contribution
\begin{equation}
F^+_1=-(4 \pi \alpha)^{1/2}\int {\frac{d^3q}{(2\pi)^3} \frac{(e q)}{m}
\frac{8\pi\alpha Z m}{(\vec p_1-\vec q)^2
(p_1^2-q^2)}\psi_F(q)},
\end{equation}
to the amplitude. In the amplitude $F^+_1$ both initial and final state
can transfer large momenta $q$ or $p_1-q$ to the nucleus. The small factor
of the order $p_1^{-4}$ is partially compensated by the large
phase volume of integration of the order $p_1^3$, multiplied by the factor
$m\alpha Z$. Thus, the amplitude $F^+_1$ is indeed $\xi_Z$ times smaller
then $F^+_0$, dropping faster with the increasing photon energy.

Thus, corrections to asymptotics provide the terms $O(\xi_Z)$,
dropping as $\omega^{1/2}$. The large numerical coefficient $\pi$
can be understood, e.g. by considering the procedure of
normalization of the function
$\psi_{p1}(r)=\psi^{(0)}_{p1}(r)+\psi^{(1)}_{p1}(r)$.  Calculating the
value $\psi_{p1}(0) =\psi_{p1}^{(0)}(0)+ \psi_{p1}^{(1)}(0)=
\int\frac{d^3q}{(2\pi)^3}
[\psi_{Fp1}^{(0)}(q)+\psi_{Fp1}^{(1)}(q)]$ one obtains:
\begin{equation}
\psi_{p1}(0)\ =\ 1+\xi_Z J
\end{equation}
with
\begin{equation}
J=\frac1\pi \int\limits_0^{\infty}
\frac{dx}{x}\ln\frac{1+x}{|1-x|}=\pi/2\ ,
\end{equation}
and thus $\psi_{p1}^{(0)}=1+\pi\xi_Z/2$. Additional power of $\pi$
can be viewed as coming from $-iln(-1)=\pi$ if the integral $J$
is evaluated in the complex plane.
Hence, the amplitude $F_0^+$ provides the asymptotics of the total amplitude.
This leads to the well known asymptotics of the cross section \cite{23}
\begin{equation}
\sigma^+(\omega)=C_1\omega^{-7/2}
\end{equation}
Because of the large corrections of the order $\pi\xi_Z$, the nonrelativistic
asymptotics can be achieved only for very light atoms. In other cases
the relativistic treatment becomes important at the energies far below the
nonrelativistic asymptotic region.

\subsection{Asymptotics of double photoionization. Spurious controversy}

The general expression for the double photoionization amplitude
can be written as
\begin{equation}
F^{2+}= \langle \Psi_f|\gamma|\Psi_i \rangle,
\end{equation}
with $\Psi_i$ and $\Psi_f$ standing for the exact wave functions of the
initial and final states, while
\begin{equation}
\gamma=\gamma_1+\gamma_2;~~~~~~~~~~~~~~ \gamma_j=
i\frac{(\vec e \vec \nabla_j)}{m}
\end{equation}
describes the interactions of the photon with the two-electron system.
The asymptotics of the amplitude \cite{20} is determined by the interactions
in the initial state, which transfer small energy $\epsilon_2 \sim I$ to the
secondary electron. Large momentum
\begin{equation}
\vec Q\ =\ \vec k-\vec p_1-\vec p_2
\end{equation}
is transferred to the nucleus.

To obtain the asymptotics, we can describe the
final state by the product of the single-particle functions, putting
$\Psi_f(\vec r_1, \vec r_2)=\psi^{(0)}_{p1}(\vec r_1)
\psi^C_{p2}(\vec r_2)$.
Here  $\psi^C$ stands for
 the continuum wave function in the Coulomb field of the nucleus,
while $\psi^{(0)}$ denotes its lowest order of expansion in powers of
$\xi_Z$ (the plane wave). Since $p_1 \gg p_2$, we can put $\gamma=
\gamma_1$.

Analysis, similar to the one, carried out in the previous subsection,
provides
\begin{equation}
F^{2+}=(4\pi\alpha)^{1/2}\frac{(\vec e \vec p_1)}{m}\int\frac{d^3q}{(2\pi)^3}
\psi^{C*}_{p2}(\vec q)\Psi_F(\vec p_1, \vec q).
\end{equation}
The integral on the rhs is saturated by small $q\sim\eta \ll p_1$. The
calculation, similar to the one presented by (25), provides
$\Psi_F(p_1,q) \sim p_1^{-4}$ at $p_1 \gg \eta$. Thus, at $p_2 \sim \eta$
the dependence of $F^{2+}$ on the photon energy is the same as that of $F^{+}$.
The integral over the energy distribution is saturated by $\epsilon_2 \sim I$
and thus does not depend on the photon energy.
Hence, the double photoionization cross section $\sigma^{2+}$ obtains the
same energy dependence as the single ionization cross section, i.e.
\begin{equation}
\sigma^{2+}(\omega)=C_2\omega^{-7/2}
\end{equation}

In paper \cite{22} the double photoionization cross section was calculated
by employing the product of the Coulomb functions for the description of the
final state. The asymptotics was obtained by expansion of the amplitude in
powers of $\xi_Z$. It was noted that the result depends on the form, assumed
for the operator of interaction between the photon and electrons. For the
{\em exact} solutions of the Schr\"odinger equation the velocity
form, presented by Eq.(33) is equivalent to the length form
\begin{equation}
\gamma=\gamma_1+\gamma_2;~~~~~~~~~~~~~~ \gamma_j=
i\omega_{fi}(\vec e \vec r_j)
\end{equation}
with $\omega_{fi}$ being the difference of the energies of the final and
initial states, $\omega_{fi}=\omega$ in our case. While the velocity form
provided the asymptotics presented by Eq.(36), the length form gave
\begin{equation}
\sigma^{2+}(\omega)\ =\ C^L_2\omega^{-5/2}.
\end{equation}
To understand this, note that since $p_2 \ll p_1$,
the important space region is  $r_1 \ll r_2$. Thus we can put
$\gamma=i\omega(\vec e \vec r_2)$. Unlike the
single photoionization, the electron, which interacts with the photon
directly, does not obtain angular momentum. The double photoionization
amplitude takes the form
\begin{equation}
F^{2+}=(4\pi\alpha)^{1/2}\omega\int\frac{d^3q_1d^3q_2}{(2\pi)^6}
\psi^{C*}_{p1}(q_1)\psi^{C*}_{p2}(q_2)(e q_2)\Psi_F(q_1,
q_2).
\end{equation}
If we describe the fast outgoing electron by the
plane wave -- Eq.(23), the dependence of the amplitude on the momentum
of the fast electron $p_1 \sim (2m\omega)^{1/2}$ is determined by the
factor
\begin{equation}
\omega\Psi_F(p_1, q_2)\ \sim\ \frac{\omega}{p_1^4}\ ,
\end{equation}
containing additional energy
dependent factor $\omega^{1/2}$, compared to Eq.(24). This leads
to Eq.(38) for the asymptotics of the cross section.
In the calculations of
\cite{22} such behavior remained true, when the fast electron was
described by the Coulomb field function.

The authors of \cite{22} suggested that the $\omega^{-5/2}$ result is
spurious, and that the corresponding contribution should vanish if the
exact wave function are used. This guess was proved by \AA berg several
years later \cite{24}.

\subsection{Solution of the controversy}

Here I present a simple derivation of the main results of \cite{24}.
Large momentum $Q \approx p_1$ can be transferred to the nucleus either by
the initial state electron, or by the final state electron. In the
length form each of these mechanisms, being treated separately,
provides the contribution of the same order in the parametric
expansion. The wave function of the outgoing electron can be described
by the superposition of the plane wave and the lowest order Coulomb
correction. Electron interactions in the final state as well as the
higher order Coulomb corrections contribute beyond the asymptotics.
 The two terms of the expansion of the final state wave function
provide the two contributions to the amplitude
$F^{2+}_L=F^{2+}_{L0}+F^{2+}_{L1}$, i.e.  \begin{equation}
F^{2+}_{Lk}=(4\pi\alpha)^{1/2}\int\frac{d^3q_2}{(2\pi)^3}
\psi^*_{Fp2}(\vec q_2)
(\vec e \vec q_2)A_k(\vec q_2)
\end{equation}
with
\begin{equation}
A_k(\vec q_2)=\omega\int\frac{d^3q_1}{(2\pi)^3}\psi_{Fp1}^{(k)*}(q_1)\Psi_F
(\vec q_1,\vec q_2)
\end{equation}
and $k=0,1$. For the plane wave contribution we find immediately
\begin{equation}
A_0(q_2)\ =\ \omega\Psi_F(p_1,q_2)\ ,
\end{equation}
with the large momentum $p_1 \gg \eta$ being transferred to the nucleus by
the initial state electrons.
(Recall that the lower index $F$ denotes the Fourier transform).
The amplitude $A_0$ is determined
by large $q_1=p_1 \gg \eta$. The integral, corresponding to $A_1$
 is saturated by small momenta $q_1 \sim \eta$. This can be
seen immediately from Eq.(27), which presents the function $\psi^{(1)}_{Fp1}$
explicitly.
The momentum  $q_1$ can be neglected everywhere, except the wave function
$\Psi_F$. This provides
\begin{equation}
A_1(\vec q_2)=-\omega\frac{8\pi\eta}{p_1^4}\int\frac{d^3q_1}{(2\pi)^3}\Psi_F
(\vec q_1,\vec q_2)=-\frac{8\pi\omega}{p_1^4}\Psi_{PF}(0,q_2).
\end{equation}
Here we introduced the partial Fourier transform
\begin{equation}
\Psi_{PF}(\vec r, \vec q_2)= \int d^3r\Psi(\vec r, \vec r_2)
e^{-i(\vec q_2 \vec r_2)}.
\end{equation}
The calculation, similar to that, presented by Eq.(25),
gives
\begin{equation}
A_0(q_2)\ =\ -\frac{8\pi\omega}{p_1^4}\Psi_{PF}^{'}(0,q_2).
\end{equation}
Thus, the amplitudes $F^{2+}_{L0}$ and $F^{2+}_{L1}$ appeared to be
of the same magnitude. (Recall, that this was not the case for the
amplitudes $F^+_0$ and $F^+_1$ due to the factor $\frac{(eq)}{m}$ on
the rhs of Eq.(28)).This does not contradict to the fact, that the
function $\psi^{(1)}(r)$ is $\xi_Z$ times smaller then $\psi^{(0)}(r)$
in any space point. The amplitudes $A_k(q_2)$, which can be presented
as
\begin{equation}
A_k(\vec q_2)\ =\ \omega\int d^3r_1\psi_{p1}^{(k)*}(\vec r_1)\Psi_{PF}
(\vec r_1,\vec q_2),
\end{equation}
are dominated by different regions of the space distances $r_1$.
While $A_0$ is determined by small $r_1\sim p^{-1}_1$, much
larger distances $r_1\sim \eta^{-1}$ provide the main contribution
to $A_1$.

Treated separately, each of the terms $A_{0,1}$ would provide
$\omega^{-5/2}$ law for the cross section. However, carrying out
the inverse Fourier transform, one finds $A_0+A_1=0$ for any
$q_2$ due to the
Kato condition (11), and thus
\begin{equation}
F^{2+}_{L0}+F^{2+}_{L1}\ =\ 0\ .
\end{equation}

This shows, how the straightforward calculations in the length form,
employing the approximate
wave functions, which do not satisfy Eq.(11), provide the spurious
$\omega^{-5/2}$ asymptotics for the double photoionization cross section.

\subsection{The Lippman-Schwinger equation}

In this subsection I show that the Lippman-Schwinger equation
(LSE) is a convenient tool for the asymptotical analysis. The general
form of LSE for the two-electron system is
\begin{equation}
\Psi\ =\ \Psi^{(0)}+G^{\epsilon}V\Psi\ .
\end{equation}
Here $V$ stands for the interaction, $\epsilon$ denotes the energy value,
corresponding to the solution $\Psi$ of the Schr\"odinger equation,
$\Psi^{(0)}$ and $G^{\epsilon}$ are the solution and the Green
function for $V=0$ (free motion).

To simplify the notations, let us write the LSE for the single-particle
wave function $\psi$, which we assume to be the exact solution of the
Schr\"odinger equation in a certain effective field $U(r)$. It is
\begin{equation}
\psi\ =\ \psi^{(0)}+G^{\epsilon}U\psi\ .
\end{equation}
Now $G^{\epsilon}$ is the free single-particle Green function.
In the momentum space
\begin{equation}
\langle \vec p |G^{\epsilon}| \vec f\rangle =
\frac{\delta(\vec p-\vec f)}{\epsilon-p^2/2m+ i\nu}~~~~\nu\rightarrow
0.  \end{equation}

For the bound states $\psi^{(0)}=0$ and
\begin{equation}
\psi_F(p)=\frac{2m}{2m\epsilon-p^2}\int{\frac{d^3q}{(2\pi)^3}
\langle \vec p |U|\vec q\rangle \psi_F(\vec q)},
\end{equation}
with $\epsilon $ standing for the single-particle binding energy. In the
asymptotics $2m\epsilon \ll p^2$.
The integral is saturated by $q \sim \eta$. Thus, Eq.(52) presents the
bound state
wave function $\psi_F(p)$ at large momenta $p \gg \eta$ in terms of the
same wave function at small momenta $q \sim \eta$.

For $s$ states we can put
$\langle \vec p |U|\vec q \rangle =
\langle \vec p |U|0 \rangle $. Since large $p$ correspond to small $r$,
only the interactions with the nucleus are important in the asymptotics.
Thus, putting
$\langle \vec p |U|\vec q \rangle=-\frac{4\pi\alpha Z}{p^2}$, we obtain
\begin{equation}
\psi_F(p)\ =\ \frac{8\pi \alpha Z\psi(0)}{p^4}+O(p^{-6}),
\end{equation}
Due to the Kato condition (15), this expression is identical to
Eq (25). In similar way we can find the asymptotics of the wave
function for the ground state of helium
\begin{equation}
\Psi_F(p,q_2)\ =\ \frac{8\pi \alpha Z\Psi_{PF}(0,q_2)}{p^4}+O(p^{-6}).
\end{equation}
with the partial Fourier transform $\Psi_{PF}$ defined by Eq.(45).

As to the wave function of the outgoing electron, the first iteration
of the LSE provides the same result, as Eq.(26) of quantum mechanics
(adding an imaginary infrared divergent term, which will be absorbed
into the phase factor, mentioned in the Introduction). Thus, calculating
the double photoionization amplitude in the length form, we immediately
find  the cancellation, expressed by Eq.(48). Hence, in the LSE technique
one would not obtain the spurious $\omega^{-5/2}$ term in the asymptotics
of the double photoionization cross section, even being not aware of the
Kato conditions.

\section{Shape of the double photoionization \newline spectrum: U or
W~?}

\subsection{Mechanisms of the process}

The progress of the experimental facilities in early 90-th made
possible the measurements of the energy distributions of the double
photoionization of helium. The low energy measurements have been
carried out. The results for the high energy are expected soon. This
stimulated the theoretical investigations.

Evaluation of the shape of the spectrum curve have been obtained in
\cite{25}. The energy distribution $d\sigma^{2+}/d\varepsilon$ reaches
its largest values at the edge of the spectrum, where
$\varepsilon_2\sim I\ll\omega$ or $\varepsilon_1\sim I\ll\omega$. The
shake-off mechanism is dominative here \cite{20}. Large momentum
$Q\gg\eta$ is exchanged between the nucleus and the fast electron.
Considering the larger values $\varepsilon_{1,2}\gg I$, we need both
electrons to carry large momenta $p_i\gg\eta$. If the energies of the
outgoing electrons are not too close, the process can be viewed as the
single photoionization, followed by the electron scattering
\cite{26,27}. This {\em final state scattering} (FSS) mechanism
dominates until we do not reach the very vicinity of the center of the
spectrum, where the {\em quasifree} (QF) process becomes possible.

To understand the point, recall that each act of the exchange by a
large momentum leads to a small factor in the amplitude. The FSS
mechanism requires the exchange by two large momenta. However in the
vicinity of the center we can manage by a single large momentum
exchange. The electron, which interacts with the photon directly, can
exchange large momentum with the second electron without participation
of the nucleus. Unlike the single photoionization, the process can go
on with the free electrons. The Bethe surface is determined by the
condition $\vec Q=0$, {\em i.e.}
\begin{equation}
\vec p_1+\vec p_2\ =\ \vec k\ ,
\end{equation}
which requires
\begin{equation}
\delta\ =\ \frac{\varepsilon_1-\varepsilon_2}\omega\ \le\
\sqrt{\frac\omega{\omega+m}}\ \ll\ 1\ .
\end{equation}
( Recall, that we do not consider very large values
of the photon energies $\omega \sim m$, which require the
relativistic treatment of the outgoing electrons.)
In other words,
the free kinematics becomes possible in the vicinity of the center of
the spectrum.

Following the general ideas, presented in the Introduction, we expect
the surplus of the spectrum curve near it's central point due to the
described QF mechanism. The energy distribution is dominated by small
values $Q\sim\eta$ of the momentum, transferred to the nucleus.

To calculate the asymptotics, we can neglect the terms of the order
$Q/p_i$ in the wave function of the outgoing electrons. The QF
amplitude can be presented as
\begin{equation}
F_{QF}(\vec p_1,\vec p_2)\ =\ F_{QF}(\vec p,\vec Q)\ =\
D(Q^2)F_{fr}(\omega,\delta) \end{equation} with $\vec p=(\vec p_1-\vec p_2)/2$.
The factor $D(Q^2)$ contains all the properties of the initial state.
Its explicit form will be given below --- Eq. (71). The amplitude
$F_{fr}$ of the absorption of the photon by two free electrons at rest
can be presented as \begin{equation} F_{fr}(\omega,\delta)\ =\ (\vec e\vec
p_1)f(\omega,\delta)+(\vec e\vec p_2)f(\omega,-\delta)\ , \end{equation} since
the electron system is a space-symmetric state. We do not need to
clarify the form of the function $f(\omega,\delta)$ here. Anyway, since
$\delta\ll1$, the lowest order of expansion in powers of $\delta$ gives
\begin{equation} F_{fr}(\omega,0)\ =\ (\vec e,\vec p_1+\vec p_2)f(\omega,0)\ =\
(\vec e\vec k)f(\omega,0)\ =\ 0, \end{equation} with the lowest order
nonvanishing term
being \begin{equation} F_{fr}(\omega,\delta)\ =\ 2(\vec e\vec p)\delta
f'(\omega,0)\ .  \end{equation} Here $f'$ denotes the derivative with respect to
$\delta$.

The energy distribution in the vicinity of the center is the interplay
of FSS and QF mechanisms. Their contributions change with the photon
energy in different ways. At $\omega\ll 2 keV$ the FSS dominates,
providing {\bf U} shaped curve. At $\omega\approx2 keV$ the QF
mechanism becomes more important. The curve obtains {\bf W} shape.

Note that usually the photoionization calculations are carried out in
the lowest order of expansion in powers of $kr_c$ with $r_c$ being the
characteristic value of the distances, which are important in the
process. This is known as the dipole approximation. In the
nonrelativistic photoionization $r_c\sim p^{-1}\sim(m\omega)^{-1/2}$ and
the dipole approximation corresponds to the lowest order of expansion
in powers of $k/p\sim(\omega/m)^{1/2}$. Going beyond the dipole
approximation have been something unusual, at least until the recent
time. The dipole approximation QF amplitude turns to zero due to Eq.
(59). This happens because the two-electron system cannot carry the
orbital momentum, equal to unity, in this kinematics. One should go
beyond the dipole approximation to obtain the nonvanishing QF
amplitude.
Unfortunately, this point was not emphasized strongly enough neither in
\cite{25}, nor in a more detailed paper \cite{27}.
The QF amplitude is expressed by Eqs. (57) and (60) with
\begin{equation}
\delta\ =\ \frac{(\vec p\vec k)}{m\omega}\ .
\end{equation}

\subsection{Dipole approximation calculations}

Meanwhile, several calculations of the high energy photoionization
spectrum have been reported. They were based on the straightforward
computation of the amplitude in the dipole approximation. The results
appeared to be controversial.  In \cite{28} the spectrum curve had
{\bf U} shaped form if the final state interactions (FSI) between the
outgoing electrons were neglected. Account of the FSI provided {\bf W}
shaped curve. The authors of \cite{29} faced the opposite situation.
Their spectrum curve for $\omega=2.8 keV$ had {\bf W} shape with a
very flat peak, when FSI have been neglected. The computations with FSI
included provided {\bf U}-shaped spectrum. Another series of
calculations without FSI \cite{30,31} gave a peak in the center of the
energy distribution.

\subsection{Explanation of the dipole approximation\newline  results}

To understand what happened \cite{32,33}, present the ground state wave
function in variables $\vec R=(\vec r_1+\vec r_2)/2$ and $\vec\rho=\vec
r_1-\vec r_2$, introducing $\tilde\Psi(\vec R,\vec \rho)=\Psi(\vec
r_1,\vec r_2)$. Describing the outgoing electrons by the plane waves,
we find for the amplitude \begin{equation} F^{(0)}=(4\pi\alpha)^{1/2}\int
d^3Rd^3\rho e^{i(\vec Q\vec R)-i(\vec p-\vec
k,\vec \rho)}\gamma\tilde\Psi(\vec R,\vec \rho).  \end{equation}
Following the analysis, carried out in Sec.3, we expect the asymptotics
$F^{(0)}(p,Q)\sim 1/(p^4Q^4)$ at $p,Q\gg\eta$. However, in the vicinity
of the center of the energy distribution we can make $Q\sim\eta$. These
values of $Q$ will be dominative in the energy distribution. The
calculation, similar to the one, presented by Eq. (25) provides
\begin{equation}
F^{(0)}=-(4\pi\alpha)^{1/2}\frac{(\vec e\vec Q)}m\cdot
\frac{8\pi\alpha}{p^4}\int d^3Re^{i(QR)}r_0
\frac{\partial\tilde\Psi(R,\rho=0)}{\partial\rho}
\end{equation}
at $p\gg\eta$, $Q\sim\eta$, with
$\partial\tilde\Psi(R,\rho=0)/\partial\rho$ denoting the partial
derivation of the function $\tilde\Psi(R,\rho)$ with respect to $\rho$
at $\rho=0$. The energy dependence of the amplitude $F^{(0)}$ is
determined by the factor $p^{-4}$, caused by the large momentum
exchange.

Large momentum can be exchanged by the outgoing electrons as well. The
lowest order perturbative correction $\psi^{(1)}$ to the final state
wave function can be obtained by using the standard formula (26) with
$V^C$ being replaced by the interaction between the electrons. The
corresponding amplitude $F^{(1)}$ is of the same order of magnitude as
$F^{(0)}$, containing the same factor $p^{-4}$. The situation is
similar to that with the shake-off amplitude in the length form, which
was discussed above. The correction $\psi^{(1)}(\rho)$ to the wave
function of the outgoing electrons is $\xi$ times smaller than the free
wave function $\psi^{(0)}(\rho)$ in each space point $\rho$. However,
the functions $\psi^{(0)}$ and $\psi^{(1)}$ contribute to the amplitude
in different space regions. While the amplitude $F^{(0)}$ is determined
by small distances $\rho\sim p^{-1}$, the amplitude $F^{(1)}$ is
dominated by the values of $\rho$ of the order of the size of the atom
$\sim \eta^{-1}$. In both cases $R\sim Q^{-1}\sim\eta^{-1}$, and both
electrons are separated from the nucleus by the distances of the order
of the size of the atom.

Direct calculation provides
\begin{equation}
F^{(1)}=(4\pi\alpha)^{1/2}\frac{(\vec e\vec Q)}m\cdot
\frac{4\pi\alpha}{p^4}\int d^3{\rm Re}^{i(QR)}\tilde\Psi(R,\rho=0)
\end{equation}
at $p\gg\eta$, $Q\sim\eta$. Each of the amplitudes $F^{(0,1)}$, treated
separately, provide the contribution, which exceeds strongly the
amplitude, describing the FSI mechanism and requiring  $Q\gg\eta$.
Hence, each of the terms $F^{(0,1)}$, treated separately, would lead to
a surplus in the energy distribution in the central region. However
\begin{eqnarray}
F^{(0)}+F^{(1)} &=& (4\pi\alpha)^{1/2}\frac{(\vec e\vec Q)}m
\frac{4\pi\alpha}{p^4}\int d^3{\rm Re}^{i(QR)}
\nonumber\\
&\times& \left(\tilde\Psi(R,\rho=0)
-2r_0\frac{\partial\tilde\Psi(R,\rho=0)}{\partial\rho}\right)=0,
\end{eqnarray}
since the last factor in the integrand turns to zero for all values of
$R$ due to Eq. (12). In the dipole approximation the leading
contribution to the energy distribution is determined by $Q\gg\eta$,
providing the ${\bf U}$-shaped spectrum curve. However, the true
distribution can be obtained only by going beyond the dipole
approximation.

Now we can explain the results of \cite{28}--\cite{31}. The authors of
\cite{28} used the ground state wave function with a weak dependence on
$\rho$. It was too weak to provide a maximum, when neglecting FSI.
Inclusion of FSI provided a spurious uncompensated ${\bf W}$ peak. In
contrast, authors of \cite{29} used wave functions which satisfied the
Kato condition (12). Thus they obtained ${\bf U}$ shape, when taking
into account the FSI. This corresponds to Eq. (65). However, a spurious
${\bf W}$ peak emerged when the FSI were neglected. The spurious central
surplus, obtained in \cite{30,31} in dipole approximation with FSI
being neglected have the same origin.

The Lippman-Schwinger equation (LSE), presented by Eq. (49) enables to
obtain these results in a simple way. The amplitude $F^{(0)}$ can be
expressed in terms of asymptotics of the Fourier transformed wave
function
\begin{equation}
F^{(0)}\ =\ -(4\pi\alpha)^{1/2}\frac{(\vec e\vec Q)}m\
\tilde\Psi_F(-{\vec Q},{\vec p})\ ,
\end{equation}
while the asymptotics of the function $\tilde\Psi_F(\vec p,\vec Q)$ is
provided by LSE equation
\begin{equation}
\tilde\Psi_F(-\vec Q,\vec p)\ =\ -\frac{4\pi m\alpha}{p^4}\int
\frac{d^3q}{(2\pi)^3}\ \tilde\Psi_F(-\vec Q,\vec q)
\end{equation}
at $Q\sim\eta$, $p\gg\eta$. The integral in the rhs is saturated by
$q\sim\eta$. Hence, the LSE presented the function
$\tilde\Psi_F(Q\sim\eta,p\gg\eta)$ in terms of this function at
$Q\sim\eta$, $q\sim\eta$. Using Eqs. (64) and (67) we immediately
obtain the cancellation $F^{(0)}+F^{(1)}=0$, without employing the Kato
condition (12).

Note that the interactions between the outgoing electrons appeared to
be more important  than their interactions with the nucleus. The latter
provide only additional corrections, while the former determine a
possible mechanism of the process. Similar analysis with the Coulomb
functions for the continuum electrons as a starting point is given in
\cite{32}.

\subsection{Necessary properties of the approximate \newline functions}

Most of the calculations of the energy spectrum are based on the
original formalism of quantum mechanics. This means straightforward
employing of the equation
\begin{equation}
d\sigma^{2+}=\ \frac\pi\omega\,|\bar F|^2\delta\left(\varepsilon_1 +
\varepsilon_2-\omega-I^{2+}\right)\frac{d^3p_1}{(2\pi)^3}
\frac{d^3p_2}{(2\pi)^3}
\end{equation}
with the amplitude $F$ calculated as
\begin{equation}
F\ =\int d^3r_1d^3r_2\psi^*_{fa}(\vec r_1,\vec r_2)\left(\gamma_1
e^{i(\vec k\vec r_1)}+\gamma_2e^{i(\vec k\vec r_2)}\right)\psi_a (\vec
r_1,\vec r_2)
\end{equation}
by the direct calculation of the integral. One should inevitably use
{\em approximate} wave functions $\psi_{fa}$ and $\psi_a$ for the
description of the final and initial states. To reproduce essential
physics, the approximate wave functions should posses certain
properties of the true wave functions.

As we have seen in Sec. 3, in the region of the shake-off domination
$\varepsilon_1\approx\omega$, $\varepsilon_2\sim I$, the FSI
correlations are not important and one can approximate the final state
by the product of the single-particle functions, putting
$\psi_{fa}(\vec r_1,\vec r_2)=\psi_{ap_1}(r_1)\psi_{ap_2}(r_2)$ and
employing the Coulomb field function for $\psi_{ap_2}$. In the velocity
form of the operator $\gamma$ one can use the plane wave for the
description of the fast electron  \cite{20}. Description of the
initial state by the product of the single-particle wave functions
\begin{equation}
\psi_a(\vec r_1,\vec r_2)\ =\ \psi_s(r_1)\psi_s(r_2)
\end{equation}
leads to the quantitatively wrong results. Thus the initial state
correlations beyond effective field approximation are important
\cite{22}.  However, there is no formula, which would enable to select
"good" wave functions. One should be even more careful, employing the
length form of the operator $\gamma$.  The lowest order Coulomb
correction to the wave function of the fast outgoing electron should be
included and Eq.  (11) should be fulfilled for the initial state
function $\psi_a(\vec r_1,\vec r_2)$ \cite{24}.

Turning to the part of the spectrum with $\varepsilon_{1,2}
\gg I$ we see that the lowest order final state interactions should be
included in the wave function $\psi_{fa}$. This is sufficient for the
proper description of all the region of FSS domination. However, to
describe the region of QF mechanism domination, one needs the wave
function $\psi_a$, for which the Kato condition (12) is true. Otherwise
the spectrum curve obtains a spurious {\bf W} peak in the dipole
approximation (the latter corresponds to replacement $e^{i(kr_{1,2})}$
by unity in Eq. (69)).

As we saw earlier, the true {\bf W} peak can be obtained beyond the
dipole approximation. This is equivalent to evaluation of the rhs of
Eq. (69), by including the second terms of the expansion
of the exponential factors $e^{i({\vec k}{\vec r}_j)}= 1+i(\vec k\vec
r_j)+\ldots$ . Such approach
provides Eq. (57) for the QF amplitude with $F_{fr}(\omega,\delta)$
defined by Eq. (60) and \begin{equation} D(Q^2)\ =\ 2r_0\int d^3R\,e^{i(QR)}
\frac{\partial\tilde\psi_a (R,\rho=0)}{\partial\rho}\ .  \end{equation} The
proper value of the derivative $\frac{\partial\tilde\psi_a
(R,\rho=0)}{\partial\rho}$ can be insured by the Kato condition (12).
Thus, even taking care of separating the quadrupole terms, one needs
the approximate function $\tilde\psi_a$, satisfying the Kato condition
(12) for the proper description of the QF central peak. Of course, any
approximate function with a nonzero value of
$\frac{\partial\tilde\psi_a (R,\rho=0)}{\partial\rho}$ provides a
central peak describing the QF mechanism qualitatively. However,
the quantitative results can be trusted only if Eq.(12)
is satisfied. Considering
such approximate function, which can be presented as combinations of
the single-particle functions (70), we find the corresponding \begin{equation}
\tilde\psi_a(\vec R,\vec \rho)\ =\ \tilde\psi_s\left(\vec
R+\frac{\vec\rho}2\right)\tilde\psi_s\left(\vec
R-\frac{\vec\rho}2\right) \end{equation} to be the even function of $\rho$. Thus,
the computations, carried out in the quadrupole approximation would
show no trace of the QF contribution. The Hartree-Fock functions
provide one of examples.

\section{Summary}

We saw that theoretical analysis, preceding the computer calculations,
appears to be very helpful. In the static problem of the calculation of
the ground state energy of helium atom it improves the convergence of
the procedure and enables to diminish the number of phenomenological
parameters. It helps to avoid "submarine rocks" in the calculations of
the asymptotics of the double photoionization cross section. It enables
also to find the true shape of the double photoionization spectrum
curve.

It was shown also that the Lippman-Schwinger equation is a powerful
tool for investigation of the asymptotics. Unfortunately,  the LSE and
the {\em Feynman diagram technique}, which can be viewed as its visual
illustration, did not yet become a part of everyday life it atomic
theory studies.

I am indebted to M.Ya. Amusia, A.I. Mikhailov, R.H.~Pratt, T.~Suri\'c
and especially to V.G.~Gorshkov for helpful discussions.


\begin{thebibliography}{99}

\bibitem{1} H. A. Bethe, in { \em Handbuch der Physik }, Springer, Berlin,
1933, v.24/1.

\bibitem{2} H. A. Bethe and R. W. Jackiw, { \em Intermediate Quantum
Mechanics}, Benjamin, NY, 1968.

\bibitem{3} M. Inokuti, Rev. Mod. Phys. {\bf 43}, 297 (1971).

\bibitem{4} R. H. Dalitz, Proc. R. Soc. A {\bf 206}, 509 (1951).

\bibitem{5} V. G. Gorshkov, JETP, {\bf 40}, 1481 (1960).

\bibitem{6} L. D. Landau and E. M. Lifshitz,
{\em Quantovaya mechanika}, Nauka, Moscow 1974-in Russian;
{\em Nonrelativistic Quantum Mechanics}, Pergamon, Oxford, 1974.

\bibitem {7} V. G. Gorshkov, A. I. Mikhailov and V. S. Polikanov,
Nucl. Phys. {\bf 55}, 273 (1964).

\bibitem {8} N. B. Avdonina, E. G. Drukarev and R. H. Pratt,
Phys. Rev. A {\bf 65}, 052705 (2002).

\bibitem {9} E. G. Drukarev and M.I. Strikman, Phys. Lett. B {\bf 186},
1 (1987).

\bibitem{10} V. G. Gorshkov, {\em Materialy VII Zimnei Shkoly LNPI}-
-in Russian, {\em Proceedings of VII LNPI Winter School}),
{\bf 2}, 415 1972.

\bibitem {11} K. S. E. Eikema, W. Ubachs, W. Vassen, and W. Hogervost,
Phys. Rev. Lett, {\bf 71}, 1690 (1993).

\bibitem{12} J. H. Barlett, J. J. Gibbons and G. G. Dunn,
Phys. Rev. {\bf 47}, 679 (1935).

\bibitem {13} V. A. Fock, Izv. Akad. Nauk SSSR,
Ser Fiz. {\bf 18}, 161 (1954)-in Russian;
K.Nor. Vidensk. Selsk. Forh. {\bf 31}, 138, 145 (1958).

\bibitem{14} M. G. Veselov and L. N. Labzovskii, {\em Teoriya Atoma. Stroenie
Electronnyh Obolochek}, Nauka, Moscow 1986-in Russian.

\bibitem{15} A. M. Ermolaev, Vest. LGU, ser. Fyz. {\bf 14}, 46 (1958);
{\bf 16}, 19 (1961). \\
A. M. Ermolaev and G. B. Sochilin, Dokl. Akad. Nauk SSSR
{\bf 155}, 1050 (1964)-in Russian; Soviet Phys.-Doklady {\bf 9}, 292 (1964).\\
 G. B. Sochilin, Int. J. Quant. Chem., {\bf 3}, 297 (1969).

\bibitem{16} C. R. Myers, C. J. Umrigar, J. P. Sethna,
and J. D. Morgan III, Phys. Rev A {\bf 44}, 5537 (1991).

\bibitem{17} T. Kato, Comm.Pure. Appl. Math. {\bf 10}, 151 (1957).

\bibitem{18} Z. Teng and R. Shakeshaft, Phys. Rev. A {\bf 47}, R3487 (1993).

\bibitem{19} M. I. Haftel and V.B. Mandelzweig, Ann. Phys. {\bf 189},
29 (1989).

\bibitem {20} P. K. Kabir and E. E. Salpiter,
Phys. Rev. {\bf 108}, 1256 (1957).

\bibitem {21} E. L. Feinberg, Dokl. Akad. Nauk SSSR  {\bf 23}, 778 (1939)-
in Russian; Journ. of Phys. {\bf 4}, 423 (1941).\\
A.B. Migdal, Journ. of Phys. {\bf 4}, 449 (1941).

\bibitem{22} F. W. Byron Jr. and C. J. Joachain,
Phys. Rev. {\bf 164}, 1 (1967).

\bibitem{23} H. Bethe, and E. E. Salpiter, {\em Quantum Mechanics of
One-and Two-Electron Atoms}, Springer-Verlag, Berlin, 1958.

\bibitem {24} T. \AA berg, Phys.Rev. A {\bf 2}, 1726 (1970).

\bibitem {25} M. Ya. Amusia, E. G. Drukarev, V. G. Gorshkov,
and M. P. Kazachkov, J. Phys. B {\bf 8}, 1248 (1975).

\bibitem {26} E. G. Drukarev, V. G. Gorshkov, A. I. Mikhailov,
and S. G. Sherman,
Phys.Lett. A {\bf 46}, 467 (1974).

\bibitem {27} E. G. Drukarev, Phys. Rev. A {\bf 52}, 3910 (1995).

\bibitem{28} M. A. Kornberg and J. E. Miraglia, Phys. Rev. A {\bf48},
3714 (1993).

\bibitem{29} Z. Teng and R. Shakeshaft, Phys. Rev. A {\bf49}, 3597
(1994).

\bibitem{30} M. A. Kornberg and J. E. Miraglia, Phys. Rev. A {\bf60},
R1743 (1999).

\bibitem{31} M. A. Kornberg and J. E. Miraglia, Eur. Phys. J. D {\bf12},
45 (2000).

\bibitem{32} T. Suri\'c, E. G. Drukarev and R. H. Pratt, Phys. Rev. A
{\bf67}, 022709 (2003).

\bibitem{33} E. G. Drukarev, N. B. Avdonina and R. H. Pratt, J. Phys. B
{\bf34}, 1 (2001).


\end{thebibliography}
\end{document}